\documentclass[aps,prl,reprint,graphicx,showpacs]{revtex4-1}
\usepackage{amsmath,amssymb,amsfonts}
\usepackage{bbold}
\usepackage{color}
\usepackage{url}
\usepackage{graphicx}

\usepackage[hyperfootnotes=true,
        pdffitwindow=true,
        plainpages=false,
        pdfpagelabels=true,
        pdfpagemode=UseOutlines,
        pdfpagelayout=SinglePage,
        hyperindex,]{hyperref}

\draft 

\newcommand{\Sdet}{\text{Sdet}}
\newcommand{\Str}{\text{Str}}
\newcommand{\tr}{\text{tr}}
\newcommand{\id}{\mathbb{1}}

\newcommand{\diag}{\text{diag}}

\newcommand{\U}{\text{U}}
\renewcommand{\epsilon}{\varepsilon}

\begin{document}

\title{Symmetry Transition Preserving Chirality in QCD: A Versatile Random Matrix Model}

\author{Takuya Kanazawa$^{1}$ and Mario Kieburg$^{2}$}
\email{mkieburg@physik.uni-bielefeld.de}
\affiliation{$^1$Research and Development Group, Hitachi, Ltd., Kokubunji, Tokyo 185-8601, Japan
\\$^2$Fakult\"at f\"ur Physik, Universit\"at Bielefeld, 33501 Bielefeld, Germany}

\newcommand{\referee}[1]{{ #1}}
\newcommand{\we}[1]{{\color{blue}#1}}

\date{\today}

\begin{abstract}
We consider a random matrix model which interpolates between the chiral Gaussian unitary ensemble and the Gaussian unitary ensemble while preserving chiral symmetry. This ensemble describes flavor symmetry breaking for staggered fermions in 3d QCD as well as in 4d QCD at high temperature or in 3d QCD at finite isospin chemical potential. Our model is an Osborn-type two-matrix model which is equivalent to the elliptic ensemble but we consider the singular value statistics rather than the complex eigenvalue statistics. We report on exact results for the partition function and the microscopic level density of the Dirac operator in the $\epsilon$-regime of QCD. We compare these analytical results with Monte Carlo simulations of the matrix model.
\end{abstract}


\maketitle 
 
\paragraph{Introduction.}

{  Random Matrix Theory (RMT) has been applied in several areas of physics. To name only a few: quantum chaos, condensed matter theory of disordered systems, and quantum information. Introductions to those and other applications can be found in~\cite{handbook:2010}. There are also applications beyond physics as in economics (finance), engineering (telecommunications), mathematics, and statistics in general. The wide applicability of RMT is referred to as \textit{universality}.
  Empirically the spectral statistics of physical systems is quite robust
  against variations in the distribution of matrix elements. The global symmetries of the system have the biggest impact to these statistics. There are just ten symmetry classes of Hermitian operators (including RMT) as proven by Altland
  and Zirnbauer~\cite{AltZirn}. When symmetries of a system are either destroyed or
  enhanced, the symmetry class usually changes continuously which has been a subject of active research for a long time~\cite{Mehta_book}. In this work we give a complete analytical solution
  to such a transition, with a strong emphasis on various applications of this result
  to high-energy physics.}

RMT was successfully applied to Quantum Chromodynamics (QCD) in the $\epsilon$-regime of QCD since the 90's~\cite{Shuryak:1992pi,Verbaarschot:1993pm,Verbaarschot:1994ip}. The reasons for the applicability of RMT in this field are that, first, QCD shares the same global symmetries with certain random matrix models and, second, the infrared eigenmodes and eigenvalues of the Dirac operator in the chirally broken phase are completely governed by these global symmetries. The two continuum theories of 3d and 4d QCD with $N_{\rm c}>2$ colours and quarks in the fundamental representation can be understood from this point of view. The corresponding random matrix model for the former is the Gaussian Unitary Ensemble (GUE)~\cite{Verbaarschot:1994ip} while the one for the latter is the chiral Gaussian Unitary Ensemble (chGUE)~\cite{Shuryak:1992pi,Verbaarschot:1993pm}.

Two main ideas have driven the RMT approach in QCD. First, the random matrix models are simple enough to provide analytical formulas relating observables like the microscopic level density of the Dirac operator to low energy constants in QCD, e.g., see~\cite{Verbaarschot:2000dy,Verbaarschot}. Those observables can then be measured with the help of lattice QCD simulations and in this way one can fix the low energy constants accurately. Secondly, RMT allows us to study situations where lattice simulations are not available, e.g., because of the sign problem. In this way it was provided a way to understand the sign problem at finite baryon chemical potential~\cite{Osborn:2004,Splittorff:2006vj,Akemann:2007rf,Kanazawa:2012zzr} and at finite $\theta$-angle~\cite{Damgaard:1999ij,Janik:1999,Kieburg:2017}.

 {  The applicability domain of RMT with chiral symmetry is not limited to QCD   
  but includes all areas where fermions with chirality emerge. 
  For example, Dirac and Weyl fermions appear in the low-energy effective
  field theories of graphene, topological insulators, semimetals and $d$-wave
  superconductors, see the reviews~\cite{dirfermrev}, which can be combined with the  RMT aproach to condensed matter and disordered systems~\cite{Beenakker}. Also non-relativistic fermions with quadratic dispersion may be described
  by chiral RMT~\cite{KanYam}.}

In the present work we wish to consider the chiral random matrix
\begin{equation}\label{RMT-model} 
\begin{split}
 \mathcal{D} =\left(\begin{array}{cc} 0 & iW \\ iW^\dagger & 0 \end{array}\right), ~
 W= H_1+i\mu H_2,\\
 H_1,H_2\in{\rm Herm}(N)\ \text{and}\ \mu\in[0,1]~~~
\end{split}
\end{equation}
distributed as
\begin{equation}\label{distribution}
P(\mathcal{D})= \frac{1}{2^N \pi^{N^2}}\exp\left[-\frac{1}{2}\tr(H_1^2+H_2^2)\right].
\end{equation}
Here ${\rm Herm}(N)$ denotes the set of Hermitian $N\times N$ matrices. 
The partition function with $N_{\rm f}$ dynamical quarks with masses $m_1,\ldots,m_{N_{\rm f}}$ is then given by
\begin{equation}\label{eq:0Zdef}
		Z_N^{(N_f)} = \int d\mathcal{D} P(\mathcal{D})\prod_{f=1}^{N_{\rm f}}\det(m_f \id_{2N} +\mathcal{D})\,.
\end{equation}
The differential $\mathcal{D}$ is the product of all real independent differentials of $H_1$ and $H_2$.
The normalization ensures that $Z_N^{(0)}=1$.  For varying $\mu$ the level statistics of $\mathcal{D}$ 
interpolates between GUE \cite{Verbaarschot:1994ip,Mehta_book} at $\mu=0$ and chGUE~\cite{Shuryak:1992pi,Verbaarschot:1993pm,Mehta_book} at $\mu=1$. Since these two limits are characterized by different patterns of flavor symmetry breaking [i.e., $\U(2N_f)\to \U(N_f)\times\U(N_f)$ at $\mu=0$ and $\U(N_f)\times\U(N_f)\to \U(N_f)$ at $\mu=1$], the interpolation is far from trivial. The random matrix $\mathcal{D}$ replaces the Euclidean Dirac operator in QCD and its matrix dimension $N$ plays the role of the spacetime volume which will be eventually taken to infinity, keeping $Nm_f^2$ and $N\mu^2$ fixed.

There are at least three major connections between the model \eqref{RMT-model} and QCD. First, \eqref{RMT-model} was proposed in~\cite{Bialas:2010hb} as a model of the 3d Dirac operator for staggered fermions. The authors of~\cite{Bialas:2010hb} employed lattice simulations to show that the microscopic spectrum of the 3d staggered Dirac operator obeys chGUE rather than GUE due to discretization effects. The possible convergence to GUE in the continuum limit can then be modeled by the model \eqref{RMT-model} with $\mu$ playing the role of the lattice spacing. The authors of~\cite{Bialas:2010hb} only compared Monte Carlo simulations of~\eqref{RMT-model} with lattice simulations, finding good agreement of the spectra. The microscopic level density of $\mathcal{D}$ has not been derived to date. {  We hope that one can estimate the lattice artefacts of the staggered discretization with the help of our analytical results.}

Another application of \eqref{RMT-model} is to 4d QCD at high temperature with twisted fermionic boundary conditions. It is well known~\cite{Ginsparg:1980ef,Nadkarni:1982kb} that gauge theories at high temperature undergoes dimensional reduction. In this regime the chiral condensate evaporates and RMT loses its validity for the infrared Dirac spectrum \cite{Kovacs:2009zj}. However, by judiciously choosing the boundary condition of quarks along $S^1$ it is possible to avoid chiral restoration up to an arbitrarily high temperature \cite{Stephanov:1996he,Bilgici:2009tx}, which has a simple explanation based on instanton-monopoles \cite{Shuryak:2012aa}. We conjecture a crossover transition of the spectral statistics of the smallest eigenvalues of the Dirac operator from chGUE (4d) to GUE (3d). {  The kind of transition will be similar if not even exactly as in our model since chirality is preserved in this situation. A comparison of our analytical results with simulations would resolve this.}

The third application of the model~\eqref{RMT-model} can be found for 3d QCD at finite isospin chemical potential. The Euclidean Lagrangian is given by
\begin{equation}\label{Lag-phys}
\begin{split}
	 \mathcal{L} = & \overline{\psi}\bigl(\mathcal{D}_{\rm 3d}
	 + \mu_{\rm I} \sigma_3\tau_3+{\rm diag}(m_{\rm u},m_{\rm d})+j\tau_1\bigl)\psi,
\end{split}
\end{equation}
where $\mathcal{D}_{\rm 3d}$ is the 3d Dirac operator which is anti-Hermitian, $\psi$ the two-flavor quark fields, $\mu_{\rm I}$ the isospin chemical potential, $m_{\rm u/d}$ the quark masses, $j$ the source term for the pion condensate, and $\tau_j$ and $\sigma_3$ the Pauli matrices in flavour space and spinor space, respectively. One can model this system by the two-matrix model \eqref{eq:0Zdef} with $N_{\rm f}=1$, in close analogy to the 4d case \cite{Osborn:2004}. (Note that $j$ in \eqref{Lag-phys} corresponds to $m_f$ in \eqref{eq:0Zdef}.) The matrix model can describe the near-zero singular values of the operator 
${\mathcal D}_{\rm 3d}+\mu_{\rm I}\sigma_3$ whose nonzero density at the origin is necessary to support nonzero pion condensate \cite{Kanazawa:2011tt}. The random matrix model for the Lagrangian~\eqref{Lag-phys} without the source $j$ was introduced in~\cite{Akemann:2001bf}. {  The application of our model to this system is a classical situation of applying RMT to QCD. The comparison of lattice simulations with our results should fix the low energy constants and thus the size of the nonzero pion condensate.}

\begin{figure*}[t!]
	\includegraphics[width=.8\textwidth]{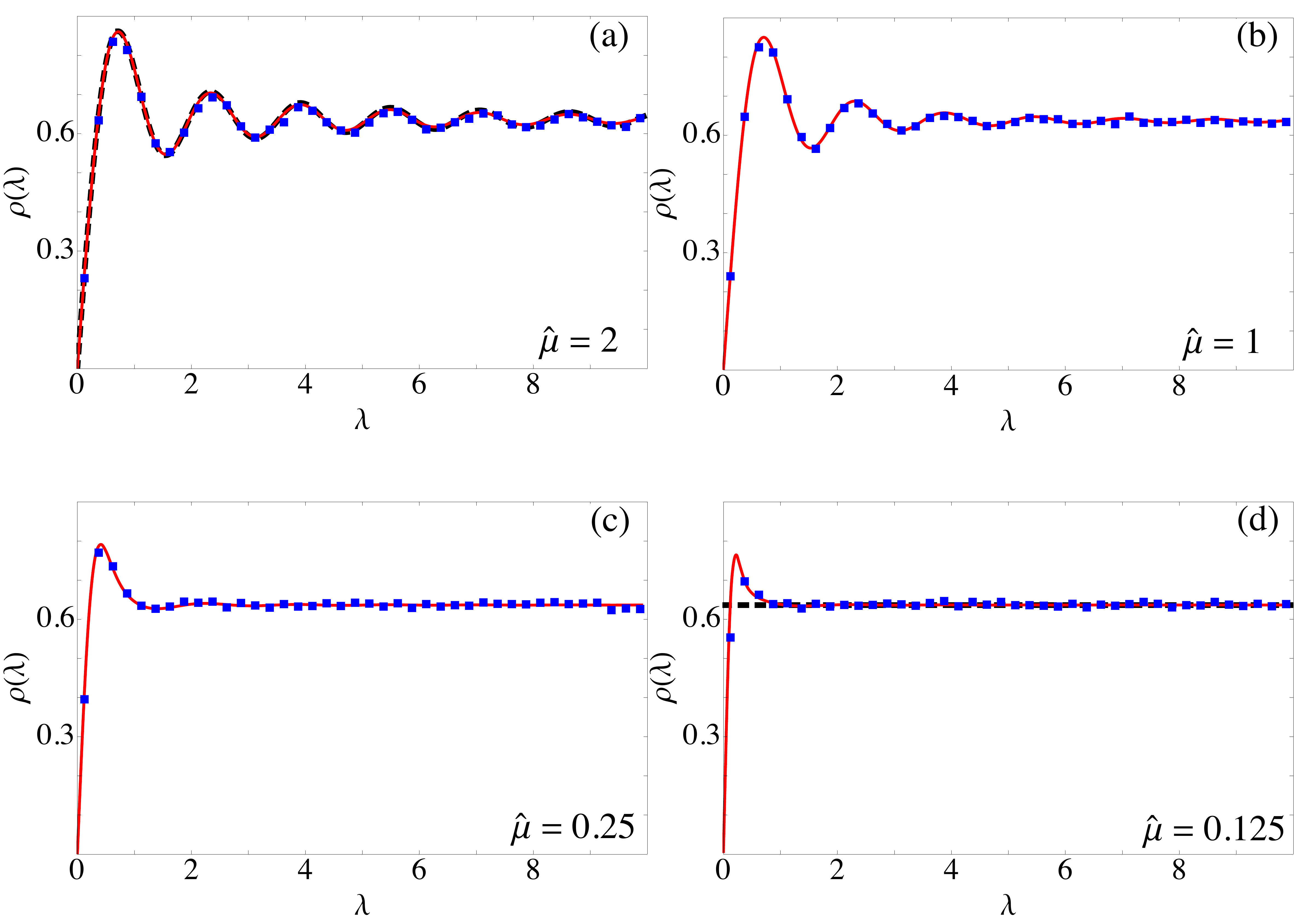}
	\caption{\label{fig.microlevel}Comparison of the analytical result~\eqref{level.density} (red solid curves) for the quenched microscopic level density with Monte Carlo simulations (blue square markers) of the random matrix model~\eqref{RMT-model} for several couplings $\widehat{\mu}$. The ensemble was generated with $10^5$ matrices and the matrix size is $n=100$ apart from plot (a) where it is $n=200$ because the effect of a finite matrix dimension is strongest there. The dashed black curves in (a) and (d) are the limiting microscopic level densities for $\widehat{\mu}\to\infty$ (chGUE) and for $\widehat{\mu}\to0$ (GUE), respectively. The normalization is given by $\lim_{\lambda\to\infty}\rho(\lambda)=2/\pi$.}
\end{figure*}

Before closing the introduction, let us briefly review preceding works that are closely related to ours. 
The random matrix $W$ is also known as the elliptic Ginibre ensemble~\cite{Fyodorov:1997,Akemann:2001bf,Akemann:2007rf} and was studied in the scattering at disordered/chaotic systems~\cite{Fyodorov:1997b} as well as in 3d QCD at finite baryon chemical potential~\cite{Akemann:2001bf,Akemann:2007rf}. The authors of these works were interested in the complex eigenvalues of $W$ while we need the singular value statistics of $W$. The singular values of $W$ are bijectively related to the eigenvalues of $\mathcal{D}$ and their statistics were not calculated before. Let us mention another random matrix model for the Hermitian Wilson Dirac operator~\cite{Damgaard:2010cz,Akemann:2011} which also describes a transition between chGUE and GUE.  The difference with our model is that the chirality of the random matrix is destroyed in~\cite{Damgaard:2010cz,Akemann:2011}. This has dramatic consequences for the statistics of those eigenvalues closest to the origin. When chirality is destroyed the level repulsion from the origin becomes weaker and weaker, see~\cite{Damgaard:2010cz,Akemann:2011}. This is not the case in our model where we preserve chirality. The microscopic level density will always drop to zero at the origin as long as $\mu\neq0$, see Fig.~\ref{fig.microlevel}.

\paragraph{Mapping to Superspace.}

To derive the chiral Lagrangian of~\eqref{RMT-model} in the $\epsilon$-regime, in particular the partition function with $N_{\rm f}$ flavors and the quenched microscopic level density, we employ the supersymmetry method~\cite{Guhr}. To this aim, we consider the partition function involving $k$ bosonic quarks and $N_{\rm f}+k$ fermionic quarks 
\begin{equation}\label{eq:Zval}
		Z_N^{(k,N_{\rm f}+k)} = \int d\mathcal{D} P(\mathcal{D})\frac{\prod_{j=1}^{N_{\rm f}+k}\det(\kappa_{{\rm f}, j} \id_{2N}+\mathcal{D})}{\prod_{j=1}^{k}\det(\kappa_{{\rm b}, j} \id_{2N}+\mathcal{D})},
\end{equation}
with $k=0$ for the ordinary fermionic partition function and $k=1$ and $N_{\rm f}=0$ for the quenched microscopic level density. The source variables are $\kappa=\diag(\kappa_{\rm b};\kappa_{\rm f})=\diag((L\epsilon-i\lambda_{\rm b})\id_k;(L\epsilon-i\lambda_{\rm f})\id_k,m_1,\ldots,m_{N_{\rm f}})$ with $m_j$ the masses of the dynamical quarks, $L=\pm1$ the sign of the regularization $\epsilon>0$ and $\lambda_{\rm b/f}$ the source variables for generating the level density. In the full version of this calculation~\cite{Kanazawa:2018b} we consider a more general partition function where $k$ and $N_{\rm f}$ are chosen arbitrarily. In the first step we rewrite the ratio of determinants in~\eqref{eq:Zval} by a Gaussian integral over an $N\times(2k|2N_{\rm f}+2k)$ rectangular supermatrix $V$. Then the integral over $H_1$ and $H_2$ is Gaussian which can be carried out explicitly, yielding
\begin{equation}\label{susy:partiton}
\begin{split}
		Z_N^{(k,N_{\rm f}+k)} = \; &\frac{1}{\pi^{2 Nk}}\int dV 
		\exp\big[\! -L\; \Str\,V^\dagger V\kappa \\
		&-\Str  (V^\dagger V\tau_1)^2/2-\mu^2\Str (V^\dagger V\tau_2)^2/2 \big].
\end{split}
\end{equation}
For the definition of the supertrace and other objects in superspace like the superdeterminant we refer the reader to~\cite{Berezin}. Note that we suppress the tensor notation with identity matrices  meaning $\kappa$ has to be understood in the $(2k|2N_{\rm f}+2k)$ dimensional superspace as $\kappa\otimes\id_2$ and the Pauli matrices $\tau_j$ as $\id_{k|N_{\rm f}+k}\otimes\tau_j$.

In the next step we apply the superbosonization formula~\cite{Zirnbauer:2006,Kieburg:2008} and replace
\begin{equation}
LV^\dagger V\tau_1\rightarrow \sqrt{N} U'=\sqrt{N}\left(\begin{array}{cc} U'_{\rm bb} & \eta^\dagger \\ \eta & U'_{\rm ff} \end{array}\right),
\end{equation}
where the the boson-boson block $U'_{\rm bb}$ satisfies $LU'_{\rm bb}\tau_1=(LU'_{\rm bb}\tau_1)^\dagger>0$ and the fermion-fermion block $U'_{\rm ff}$ is unitary. The $2k\times(2k|2N_{\rm f}+2k)$ off-diagonal block $\eta$ comprises only independent complex Grassmann variables. In terms of this supermatrix, the partition function is up to a global constant equal to
\begin{equation}\label{susy:partiton.b}
\begin{split}
		Z_N^{(k_{\rm b},k_{\rm f})} \propto&
		\int d\tilde{\mu}(U')\; \Sdet^{N}U' \exp\big[\!-\Str\,U'\widehat{\kappa}\tau_1
		\\
		& - N\, \Str\,{U'}^2/2 + \widehat{\mu}^2\Str (U'\tau_3)^2/2 \big].
\end{split}
\end{equation}
Here we have already chosen the rescaling from the $\epsilon$-regime, namely $\widehat{\kappa}=\sqrt{N}\kappa$ and $\widehat{\mu}^2=N\mu^2$ which are fixed in the large-$N$ limit.

\paragraph{Chiral Lagrangian.}

First we want to consider the partition function with $N_{\rm f}$ dynamical quarks, i.e. $k=0$ \cite{comm}. There is no supersymmetry and the whole supermatrix consists of the fermion-fermion block, only, $U'=U_{\rm ff}$. For $N\to\infty$ we have to solve the saddle point equation
\begin{equation}\label{sad.eq}
 U'^2=\id_{2N_{\rm f}}.
\end{equation}
This implies that all eigenvalues of $U'$ are $\pm1$. When diagonalizing $U'$ to its eigenvalues $e^{i \phi}=\diag(e^{i \phi_1},\ldots,e^{i \phi_{N_{\rm f}}})$ we would obtain the Vandermonde term $|\Delta_{N_{\rm f}}(e^{i \phi})|^2=\prod_{a<b}|e^{i \phi_b}-e^{i \phi_a}|^2$ as the Jacobian. This term algebraically suppresses all solutions of~\eqref{sad.eq} in $1/N$ which do not have an equal number of $\pm1$. Thus we integrate only over the coset ${\rm U}(2N_{\rm f})/[{\rm U}(N_{\rm f})\times {\rm U}(N_{\rm f})]$ which is the same as for $\mu=0$ (GUE, 3d QCD), see~\cite{Verbaarschot:1994ip}. This coset can be parametrized as $U'=U\tau_3 U^\dagger$ with $U\in{\rm U}(2N_{\rm f})$. The chiral Lagrangian contains a perturbation to the $\widehat{\mu}=0$ result~\cite{Verbaarschot:1994ip,Szabo:2005gi} and the partition function in the large-$N$ limit is given as
\begin{align}\label{k-flav-f}
Z^{(0,N_{\rm f})}=\!\!\!\!\! \int\limits_{\U(2N_{\rm f})} \!\!\!\!\!\!
d\mu(U)\exp\left[\tr\,U\tau_3U^\dagger \widehat{\kappa}\tau_1-\frac{\widehat{\mu}^2}{2}\tr(U\tau_3U^\dagger\tau_3)^2\right]\!.
\end{align}
The symmetry crossover is controlled by the second term. It gives nonzero masses to splitt off the Nambu-Goldstone modes, reducing the coset down to $\U(N_{\rm f})$ for large $\widehat{\mu}$. We expect that the result \eqref{k-flav-f} reproduces the leading order of the $\epsilon$-expansion in QCD for the three areas of applications discussed above. In~\cite{Kanazawa:2018b} we have even carried out the integral over $U$ explicitly, which yielded either an $N_{\rm f}\times N_{\rm f}$ or an $(N_{\rm f}+1)\times(N_{\rm f}+1)$ Pfaffian determinant, depending on whether $N_{\rm f}$ is even or odd, respectively. This Pfaffian structure was also obtained in finite $N$ computations by us in~\cite{Kanazawa:2018a}.

\paragraph{Microscopic Level Density.}

Next we want to derive the microscopic level density in the quenched limit. Its relation to the partially quenched partition function for $k=1$ and $N_{\rm f}=0$ is
\begin{equation}\label{level-part}
\rho(\lambda)=\frac{1}{\pi}\lim_{\epsilon\to0}{\rm Re}\left.\partial_{\widehat{\kappa}_{\rm f}}Z_N^{(1,1)}\right|_{\widehat{\kappa}_{\rm b}=\widehat{\kappa}_{\rm f}=L\epsilon-i \lambda}.
\end{equation}
In the large-$N$ limit the saddle point equation~\eqref{sad.eq} still applies though with a different dimension in the unit matrix. For this reason we use the parametrization
\begin{equation}
\begin{split}
{U'}_{\rm bb}& = Le^{-\vartheta_1\tau_3/2}e^{-\vartheta_2\tau_2/2}\tau_1e^{z_1\id_2+z_2 \tau_1}e^{\vartheta_2\tau_2/2}e^{\vartheta_1\tau_3/2},\\
{U'}_{\rm ff}& = e^{-i\varphi_1\tau_3/2}e^{-i\varphi_2\tau_2/2}\tau_1e^{i(z_3\id_2+z_4 \tau_1)}e^{i\varphi_2\tau_2/2}e^{i\varphi_1\tau_3/2},
\end{split}
\end{equation}
for the two diagonal blocks of $U'$. The angles take the values $z_1,z_2,\vartheta_1,\vartheta_2\in\mathbb{R}$, $\varphi_1\in[-\pi,\pi]$, and $\varphi_2,z_3,z_4\in[-\pi/2,\pi/2]$. From the saddle point equation and the properties of ${U'}_{\rm bb}$ and ${U'}_{\rm ff}$ we have to expand the variables $z_j$ as follows: $(z_1,z_2)=(\delta z_1/\sqrt{N},\delta z_2/\sqrt{N})$ and $(z_3,z_4)=(\delta z_3/\sqrt{N},$ $\delta z_4/\sqrt{N}), ~(z_3,z_4)=(\pi/2+\delta z_3/\sqrt{N},\pi/2+\delta z_4/\sqrt{N})$. The latter expansion for $z_3,z_4$ yields an algebraically $1/N$-suppressed term. The expansion to leading order in $1/N$ and integration over $\eta$ and $\delta z_j$ is quite lengthy and we omit it here. It is done in detail in~\cite{Kanazawa:2018b}. Afterwards we integrate over $\varphi_1$ and $\vartheta_1$ which yields the modified Bessel functions of first order, $I_j$, and of second order, $K_j$, respectively. Then we are able to use the relation~\eqref{level-part}. One important identity for the Bessel functions is the following~\cite[Eq.~(9.6.1)]{Abramowitz},
\begin{equation}
\lim_{\epsilon\to0}{\rm Im}(\epsilon-i \lambda)^\nu K_\nu(2(\epsilon-i \lambda)\cosh\vartheta_2)=\frac{\pi}{2}\lambda^\nu J_\nu(2\lambda\cosh\vartheta_2)
\end{equation}
which simplifies the result a lot. Substituting $x=\sin\varphi_2$ and $y=\sinh\vartheta_2$, we eventually obtain an expression for the microscopic level density of the quenched system,
\begin{widetext}
\begin{align}
\rho(\lambda)=\;&\frac{2}{\pi}\int_{-1}^1 dx\int_{-\infty}^\infty dy\exp\left[-2\widehat{\mu}^2\left(x^2+y^2\right)\right]\biggl[\left(2\widehat{\mu}^4(y^2-x^2+1)(y^2-x^2)-\widehat{\mu}^2\frac{x^2+y^2}{2}-\frac{1}{2}\lambda^2(2+y^2-x^2)-\frac{1}{4}\right)\notag\\
&\times\sqrt{1-x^2} J_1(2|\lambda|\sqrt{1-x^2})J_0(2\lambda\sqrt{1+y^2})+\lambda^2(1-x^2)\sqrt{1+y^2}J_0(2\lambda\sqrt{1-x^2})J_1(2|\lambda|\sqrt{1+y^2})\notag\\
&+\left(2\widehat{\mu}^2(y^2-x^2+1)+\frac{1}{4}\right)|\lambda|(1-x^2)J_0(2\lambda\sqrt{1-x^2})J_0(2\lambda\sqrt{1+y^2})\notag\\
&+\left(2\widehat{\mu}^2(y^2-x^2+1)+\frac{1}{4}\right)|\lambda|\sqrt{1+y^2}\sqrt{1-x^2}J_1(2\lambda\sqrt{1-x^2})J_1(2\lambda\sqrt{1+y^2})\biggl]\,.
\label{level.density}
\end{align}
\end{widetext}
This is a new result. We normalized it to the asymptotics $\lim_{\lambda\to\infty}\rho(\lambda)=2/\pi$.
This density is plotted in Fig.~\ref{fig.microlevel} and compared with Monte Carlo simulations of the random matrix model~\eqref{RMT-model} for several $\widehat{\mu}$. One immediately notices the different behaviour of this result compared to the one in~\cite{Damgaard:2010cz,Akemann:2011}, where the transition between chGUE and GUE was done in a way which breaks chirality, reflecting the form of the Wilson Dirac operator. Due to the preservation of chirality in our model, the GUE limit ($\widehat{\mu}\to0$) is not uniform about the origin while it is in~\cite{Damgaard:2010cz,Akemann:2011} when the quark mass vanishes. As we have already pointed out, an exact chirality yields for a complex matrix a linear level repulsion from the origin, though the range where this is happening is of order $\mathcal{O}(\widehat{\mu})$ for small $\widehat{\mu}$. However the two models also have something in common. The oscillations are increasingly suppressed when $\widehat{\mu}$ decreases, whereas the limit $\widehat{\mu}\to\infty$ to the chGUE result is uniformly approached.

\paragraph{Conclusions.}

We sketched the derivation of the chiral Lagrangian, see Eq.~\eqref{k-flav-f}, and the microscopic level density, see Eq.~\eqref{level.density}, in the $\epsilon$-regime of QCD for the random matrix model~\eqref{RMT-model} which interpolates between chGUE ($=$ 4d continuum QCD) and GUE ($=$ 3d continuum QCD). The details of these computations are done in~\cite{Kanazawa:2018b} and finite $N$ results are given in~\cite{Kanazawa:2018a}. Again, we underline that the crucial difference with the related models in~\cite{Damgaard:2010cz,Akemann:2011} is the preservation of chirality which has drastic consequences on the eigenvalues close to the origin. However we do not expect any differences for the spectral statistics in the bulk and at the soft edges. The infra-red physics of QCD in three and four dimensions, as we mentioned earlier, is directly related to the Dirac eigenvalues closest to the origin and we believe our results can be used to quantify effects like the discretization via staggered fermions in 3d or of a high temperature in 4d {  with twisted boundary conditions. The quantification should work in the standard way by comparison of our analytical results with lattice simulations.}

The results we presented here can be readily generalized to the situation of $N_{\rm f}$ dynamical quarks as well as to higher order correlation functions. In~\cite{Kanazawa:2018a} we have shown that the spectral statistics describe a Pfaffian point process meaning that all $k$-point correlation functions as well as averages over ratios of characteristic polynomials of the random matrix $\mathcal{D}$ are given by Pfaffian determinants which only comprise three functions. This simplifies the analysis greatly and it carries over from finite matrix dimension $N$ to infinite matrix size. In particular the statistics of the infra-red eigenvalues of the physical Dirac operators in QCD (say 3d staggered Dirac operators or 3d continuum theory with isospin chemical potential) should also follow this Pfaffian point process and all the consequences related to it.

{  Given the connection of RMT to various fields of physics we anticipate that
   the solution presented in the present work would be of value outside of QCD as well,
   especially in the field of condensed matter systems with emergent Dirac
   fermions.}

\paragraph{Acknowledgements.}
We acknowledge support by the RIKEN iTHES project (TK) and  the German research council (DFG) via the CRC 1283: ``Taming uncertainty and profiting from randomness and low regularity in analysis, stochastics and their applications" (MK).


\begin{thebibliography}{10}
\bibitem{comm}
We count the number of four-component Dirac fermions by $N_{\rm f}$. In 3d, it is equivalent to $2N_{\rm f}$ two-component Dirac fermions.
\end{thebibliography}
\end{document}